\documentstyle[12pt]{article}
\newcommand{\dia}{\begin{displaymath}}
\newcommand{\die}{\end{displaymath}}
\newcommand{\eqa}{\begin{equation}}
\newcommand{\beq}{\eqa}
\newcommand{\eqe}{\end{equation}}
\newcommand{\eeq}{\eqe}
\newcommand{\eqna}{\begin{eqnarray}}
\newcommand{\beqa}{\eqna}
\newcommand{\eqne}{\end{eqnarray}}
\newcommand{\eeqa}{\end{eqnarray}}
\newcommand{\eqnaa}{\begin{eqnarray*}}

\newcommand{\eqnae}{\end{eqnarray*}}

\renewcommand{\Re}{\mbox{\rm Re}}
\renewcommand{\Im}{\mbox{\rm Im}}
\title{Trapping of Projectiles in Fixed Scatterer Calculations\footnote{
  This work was supported in part by the Bundesministerium f\"ur
  Forschung und Technologie  (SL) and by the Japan Society for
  the Promotion of Science (DS)}
}
\author{S.Lenz\\ Institut f\"ur theoretische Physik III, Universit\"at
Erlangen, \\ Staudstr.7 \\ 91058 Erlangen, Germany\and D.Stoll\\
Department of Physics,
University of Tokyo, Hongo 7-3-1, \\ Bunkyo-ku, Tokyo 113, Japan}
\begin{document}
\maketitle
\begin{abstract}
   We study multiple scattering off nuclei
   in the closure approximation. Instead of reducing the dynamics to
   one particle potential scattering, the scattering amplitude
   for fixed target configurations is
   averaged over the target groundstate density via stochastic
   integration. At low energies a strong coupling limit is found which
   can not be obtained in a first order optical potential approximation.
   As its physical explanation, we propose it to be caused by
   trapping of the projectile. We analyse this
   phenomenon in mean field and random potential approximations.
   (PACS: 24.10.-i)
\end{abstract}

\section{Introduction}
Multiple scattering off composite targets has been a field of interest
for the last decades in several branches of physics. Whenever the
interaction between target and projectile is strong or the
target system is dense, calculations which take into account multiple
scattering by truncating the Born series fail. Unless very specific reactions
are considered, physics in this regime will be dominated by genuine
multiple scattering effects which is particularly the case for
elastic scattering. Although for elastic scattering it may
be proven that the many body problem is equivalent to a one body
problem with a nonlocal potential \cite{Foldy}, the explicit construction
of this optical potential requires solving a many body problem involving
the target degrees of freedom. Approximate solutions are obtained
by assuming that only a small number of target states dominate the dynamics.
In the first order optical potential approximation target propagation
is restricted to its groundstate, in which case the optical potential
for s--wave projectile--nucleon interaction
becomes local with complex strength. Despite its simplicity,
the first order optical potential has been applied
successfully to many reactions in nuclear physics. More recently it has been
used in the analysis of low energy scattering of $\rm\bar p$ and
$\rm K^-$--mesons off nuclei and the corresponding atomic systems
(for a review and references see \cite{Batty}).
In these cases, however, it is found that a description
of the experimental data is not possible using the free space
t--matrices extracted from projectile nucleon scattering data.
Due to the large $\rm \bar p p$ and $\rm K^- p $ scattering
amplitudes the projectile wavefunction calculated in the first order optical
model is suppressed in the nuclear interior, thus giving rise to
scattering phase shifts close to hard sphere scattering. Phenomenological
description  of the data
requires a change in the real part of the potential from repulsive
to attractive which has been attributed to missing
dynamical properties in a potential model \cite{Friedman}.

In this work we do not investigate a specific reaction process.
Rather we want to demonstrate that, within a theoretical model study,
there exists a so far unobserved multiple
scattering phenomenon which is not found when reducing the complex
scattering dynamics to potential scattering. To this end
we contrast the first order optical potential with another
approach, in which all excited nuclear states contribute as intermediate
states. The underlying approximation which is the
so called fixed scatterer or frozen nucleus approximation leads to a
dynamically richer model in which projectile scattering off the
nucleus can even for s--wave interaction not be reduced to a scattering
problem with a local potential. The scattering amplitude in this approach
is calculated for a fixed spatial configuration of nucleons first and
then averaged over the nuclear groundstate density (for an application
of this method to $\pi$--scattering see \cite{Gibbs,Bethe}).
For large projectile nucleon scattering amplitude no hard sphere phaseshifts
like for the first order optical potential are found.
The projectile nucleus scattering amplitude becomes independent
of the elementary scattering amplitude and is strongly imaginary.
This strong coupling limit is related to the generation of a new
lengthscale by averaging over resonant individual
configurations. We propose trapping of the projectile in the target
system as an explanation.

The paper is organized as follows: In section {\bf \ref{Fixed}} we briefly
review the formal multiple scattering theory and the derivation of our model.
We concentrate on zero energy scattering in section {\bf \ref{Res}}.
Section {\bf\ref{incl}}
presents a study of inelastic cross sections
in which the physical interpretation
of the strong coupling limit becomes more transparent. Section
{\bf \ref{Rand}} is
devoted to the discussion of random potential models which will be found
to be insufficient for reproducing the strong coupling limit.
We end up with a summary
and concluding remarks.

\section{Derivation of the models}
\label{Fixed}
As the formal theory of multiple scattering may be found in many textbooks
\cite{Rod,Gold}, we discuss this topic only briefly in order to be able to
state the approximations underlying the fixed scatterer and
the first order optical potential model.
We start from a Hamiltonian of the form
\beq
   H = H_0 + K + \sum_{i=1}^A v_i \quad ,
\eeq
describing a system of $A$ target particles whose motion is
governed by $H_0$ and a projectile with kinetic energy $K$
interacting with the $i$-th target particle via a twobody
potential $v_i$.
The usual way of formally solving this problem is to
introduce projectile-nucleon transition operators $\tau_i$
\beq
   \tau_i = v_i + v_i G \tau_i ,
   \label{tau}
\eeq
where $G$ is the Green's function for the noninteracting projectile target
system
\beq
   G = [ E - H_0 - K + i \epsilon]^{-1}
     = \sum_n  |n\rangle\langle n | [ E - E_n - K + i\epsilon]^{-1}
\eeq
with outgoing wave boundary conditions. $|n \rangle$ is
a complete set of nuclear energy eigenstates of $H_0$
with energies $E_n$.
The transition matrix describing the scattering of the projectile
by the many body target is found to be
the solution of the following set of linear operator equations
\cite{Gold}
\beqa
    T & = & \sum_i T_i \\
    T_i & = & \tau_i + \tau_i G \sum_{i \neq j} T_j
    \label{Ti} \quad .
\eeqa
The transition operators $\tau_i$ are the solution of the scattering problem
if only one potential $v_i$ is different from zero. As their
calculation involves $G$ they still contain all the
target dynamics and are in general many body operators.
Therefore a solution may only be obtained after
approximating this set of operator equations.
One approximation has to be made on the level of the projectile
nucleon interaction. We assume
that $\tau_i$ may be replaced by the free space projectile-nucleon
transition matrix $t_i$. This so called impulse approximation is valid
if the projectile-target interaction time is much shorter than
the typical time scale of the target.
The impulse approximation is common to the scatterer and
first order optical potential model.
To obtain a solvable model, an additional approximation on the
intermediate states between two scattering events is needed.
The fixed scatterer approximation consist of replacing
the Green's function in eq.(\ref{Ti})
by the so called closure Green's function
\beq
   G_0 \approx G_{closure} = [E- E_0-K + i \epsilon]^{-1} ,
\eeq
which means that we neglect the motion of the target
particles while the projectile travels inside the target. Note that
both these assumptions become justified if the mass
difference between target particles and projectile is very large.
The consequence of the fixed scatterer
approximation is that the projectile cannot loose energy
while propagating inside the target system. Although the target
particles remain fixed in a scattering event, the many body aspect is
still present since quantum mechanical scattering amplitudes are obtained
only after averaging over different sets of configurations of the
target particles. For an excellent discussion
of this approximation see \cite{Foldy}.
We shall see that even after reducing the target dynamics to a
mere averaging process the model is still capable of producing
nontrivial phenomena.

If the target particles do not overlap and the nucleon projectile interaction
is s-wave dominated the scattering amplitude of a configuration of target
centers at positions $\{ \vec{r}_i \}$ is given by \cite{Foldy,Can}
\beqa
   F_{\vec k^{\,\prime},\vec k}(\vec r_1,\dots ,\vec r_A)&=&
   f_0 \sum_i \mbox{e}^{-i\vec k^{\,\prime}\vec r_i}\psi_i \\
   \psi_i&=&\mbox{e}^{i\vec k \vec r_i} +f_0 \sum_{j\ne i}^{A}\frac{
   \mbox{e}^{ik|\vec r_j-\vec r_i|}}{|\vec r_j-\vec r_i|}\psi_j
   \label{a26}               ,
\eeqa
where $f_0$ is the elementary projectile-nucleon s-wave scattering
amplitude taken from \\ $\langle\vec k^{\,\prime}|t_i|k\rangle $ which is
assumed to be equal for all target particles.
The full elastic  scattering amplitude is obtained
by averaging the individual amplitudes with the weight given by the
target ground state density
$\rho_0=|\langle\vec r_1,\dots,\vec r_A|0\rangle|^2 $:
\eqa  F(\vec k^{\,\prime},\vec k)=
   \langle F_{\vec k^{\,\prime},\vec k}(\vec r_1,\dots ,\vec r_A)\rangle
   :=\int d^3r_1\dots d^3r_A\rho_0 (\vec r_1,\dots
  ,\vec r_A) F_{\vec k^{\,\prime},\vec k}(\vec r_1,\dots ,\vec r_A).
  \label{a27}
\eqe
Due to its high--dimensional nature
we solve the integral (\ref{a27}) by stochastic integration (for
details of the method see \cite{Lenz,Kalos}).
$N$ configurations of nucleon postitions $\{ \vec{r}_i^{(n)} \}$ are
sampled from the groundstate density and the integral is then
approximated by
\beq
   \langle  F_{\vec k^{\,\prime},\vec k} \rangle  \approx \langle
   F_{\vec k^{\,\prime},\vec k}\rangle _N
   := \frac{1}{N} \sum_{n=1}^{N}
   F_{\vec k^{\,\prime},\vec k}(\vec r_1^{(n)},\dots ,\vec r_A^{(n)})
   .
\eeq
As in the limit $N \rightarrow \infty$ the sum
$\langle F_{\vec k^{\,\prime},\vec k}\rangle _N$ approaches
the exact value of the integral, the elastic scattering amplitude can
be calculated in principle to arbitrary precision.

The first order optical potential which will be used for comparing
the results of our model is constructed from eq.(\ref{Ti})
by approximating the Green's function assuming that the
nucleus will stay in the ground state between successive scatterings
with the projectile
\[
   G_{opt.} \approx | 0 \rangle
   \langle  0 | [ E -E_0 - K + i \epsilon ]^{-1}
{}.
\]
If one further drops the summation restriction $(i\ne j)$
in eq.(\ref{Ti}), the elastic transition matrix for the projectile
$t = \langle 0| T |0\rangle $ can
be calculated from a one-body equation.
\[
  t = U + U [E-E_0 - K + i \epsilon]^{-1} t
\]
with the first order optical potential
\[
 U = \sum_i \langle 0| t_i | 0\rangle
 .
\]
The averaging over the nucleon positions is
performed in this model already on the level of the projectile-nucleon
scattering operators. This is in strong contrast to the
fixed scatterer model discussed before, where this averaging is done
at the latest stage after multiple scattering has been calculated
to infinite order. The consequence of the different
treatment of intermediate states can be clearly seen from an expansion
of the Born series of both models.
In the fixed scatterer model diagrams are found in which the projectile
returns to a scattering center after visiting another.
These diagrams are
are absent in the first order optical potential \cite{Foldy,Bethe}.

The importance of backscattering correlations is well known
e.g. in the tight binding model description of disordered
lattices \cite{Anderson,Souillard,Ziman}.
Anderson localization is found only if certain classes of
backscattering diagrams are taken into account to infinite order.
Coherent potential approximations which are the tight binding model analog
of the first order optical potential approximation are not able to
describe the metal-insulator transition the Anderson model predicts
(see \cite{Ram} for a recent review). The analysis of the tight binding model
in general makes use of the fact that a lattice site is only coupled
to a small number of neighbouring sites.
Formally this corresponds to the situation where the sum in equation
(\ref{a26}) extends only over few centers. Under such conditions it was
shown \cite{Anderson} that iteration
of a small number of elementary diagrams is sufficient to understand
the physics contained in backscattering correlations.

Similar approaches have been used in nuclear physics in order to construct
improved optical potentials \cite{Keister}.
The results to be discussed below, show that at low energy
and for large elementary scattering amplitude there exists
a strong coupling limit, which is related to backscattering
correlations of at least four nucleons.
Therefore improved optical potential may in this limit
still miss important physics.
This seems to be plausible because of the formal differences
between the multiple scattering equation (\ref{a26})
and the corresponding equation in the tight binding model.
In the latter the assumption of only nearest neighbour interaction
is frequently used, in contrast to equation (\ref{a26}), where
the coupling between two centers decreases only like the inverse of
their distance. Therefore an expansion of
(\ref{a26}) in terms which only involve the positions of part of the
scattering centers will fail as soon as $f_0$ is larger than the
typical internucleon distance.
In this limit configurations of nucleons which trap the projectile
by multiple scattering cause a completely different result as
compared to the first order optical potential.

The qualitatively different behavior of the results in the first order optical
potential model and the fixed scatterer model can be most clearly seen
in the zero energy limit where an understanding can be gained
with a minimum of formal tools.
Therefore we first concentrate on
this case.

\section{Zero energy scattering}
\label{Res}
In this section we demonstrate the existence of a strong coupling
limit in the fixed scatterer model, we study the target particle number
dependence  and we compare
the results of the fixed scatterer model
with first order optical potential calculations.
First the results for scattering off $^4\rm He$ and
$^{16} \rm O$ are discussed.
We used harmonic oscillator densities with
size parameter $b=1.41 \rm \, fm$ for $^4 \rm He$ and $b=1.71 \rm \, fm$
for $^{16} \rm O$.
These values were chosen to fit the nuclear rms--radius \cite{Lenz,Jaeger}.
Figure \ref{Real4} shows real and imaginary part of the forward
scattering amplitude $F ( \theta=0 )$ for $^4 \rm He$
as function of $|f_0|$ for three different
arguments of $f_0$. For all calculations
$N = 2 \times 10^4 $ configuration were used.
We clearly observe that in the region where $|f_0|$ is comparable with the
typical internucleon distance the results still depend strongly
on the phase of $f_0$. However, for larger $|f_0|$ a saturation value
$F(\theta=0)\approx -2.2{\rm \, fm} + i \times 0.6{\rm \, fm}$ is reached.
A comparison
between the optical potential and the fixed scatterer calculation is
shown in figure \ref{O16} where the ratio of imaginary to real part
of the scattering amplitude off $ ^{16}\rm O$ is plotted. The stochastic
integration was performed with $N=10^4$ configurations.
As we can see, first order optical potential and stochastic calculation are
still in qualitative
agreement for not too large elementary scattering amplitude. In the
limit of large elementary amplitude the  stochastic result reaches
a phase independent limit of $\Im F / \Re F \approx -0.55 $, whereas
the optical potential calculation yields a strongly phase dependent
and decreasing result which is easily understood
from the fact that the wavefunction vanishes inside potentials
with large negative imaginary part. Therefore the optical potential
predicts that the projectile can not penetrate deeply into the nucleus
but is scattered in a surface region.
The phase dependence is due to to the change of the real part of the
potential from attractive to repulsive. For a repulsive real part the
surface region becomes smaller than it is for an attractive one,
which leads to a smaller ratio of imaginary to real part of the scattering
amplitude.

The universal limit in the fixed scatterer calculation is
reached at values which are much larger than scattering lengths which
are found experimentally in strongly interacting systems as $\rm\bar pp$.
Nevertheless from figure \ref{O16} it may be deduced that precursors of
this limit appear much earlier.  To find the reason for the different
behavior we study in the following
our model in the limit $|f_0| \rightarrow \infty$.

In order to find the origins for the constancy of the results
in the fixed scatterer model we formulate a simple mean field model.
We replace the $k=0$ propagator in eq.(\ref{a26}) by its average
\[
   \frac{1}{| \vec{r}_i - \vec{r}_j| } \approx \frac{1}{R_0} =
   \langle  \frac{1}{| \vec{r}_i - \vec{r}_j| } \rangle  ,
\]
which means that we drop all the fluctuations in the configuration
ensemble.
With this approximation the system of equations is trivial and the
scattering amplitude becomes
\beq
    \langle F\rangle _{mf} = \frac{ A f_0 }{ 1 - (A-1) f_0 / R_0 }
       \stackrel{f_0 \rightarrow \infty}{=} - \frac{A}{A-1} R_0
{}.
\label{mean}
\eeq
For not too large elementary scattering amplitude
the mean field model describes  real and imaginary part of the
scattering amplitude rather well but in the limit
$|f_0| \rightarrow \infty$ it predicts a zero imaginary part which
will be discussed later.
To test the $A$ dependence of the scattering amplitude in this
approximation we placed $A$ centers in a homogenous density of the form
\[
    \rho^{(A)} ( \vec{r}_1 , .. , \vec{r}_A ) =
    \rho_0 \prod_{i=1}^{A} \Theta (R - | \vec{r}_i | )|
\]
with $R=2.6\rm \, fm$, leading to a value $R_0\approx 2.2\rm \, fm$.
Figure \ref{Adependence} shows a comparison of the mean
field model and the exact fixed scatterer results
for $f_0 = 250.0{\rm \, fm} + i \times 6.3 {\rm \, fm}$.
The scattering amplitude was choosen
on the one hand to be much larger than $R_0/A$,
which is the relevant scale
for the set in of universality. On the other hand the imaginary part
was taken small compared to the
real part to demonstrate that our result of the previous
examples is not caused by large elementary imaginary parts.
However, due to the almost real value of $f_0^{-1}$,
$N = 4 \times 10^6 $ configurations
were needed to ensure stochastic convergence.

We observe that for a large enough number of
scattering centers the mean field
prediction for the real part of the scattering amplitude
is in excellent agreement with the stochastic calculation.
The imaginary part of the scattering amplitude
is almost zero for 2 and 3 centers whereas for 4 and more centers it reaches
a nonzero value which is only weakly dependent on $A$ from $A=8$ on.
To understand the origin of the imaginary part of the scattering amplitude
we rewrite eq.(\ref{a26}) in the following way
\beq
    F =  \sum_{ij} (\frac{1}{ \frac{1}{f_0} - G} )_{ij} ,\quad
    G_{ij} = \frac{1}{| \vec{r}_i - \vec{r}_j |}  (1-\delta_{ij})
    .  \label{gij}
\eeq
If one inserts a complete set of eigenvectors of $G$ defined by
\[
   \sum_j G_{ij} \Phi^{(n)}_{j} = \lambda_n \Phi^{(n)}_i ,
\]
the scattering amplitude reads
\beq
   F = \sum_n \frac{r_n^2}{\Re\frac{1}{f_0} + i \Im \frac{1}{f_0} -
       \lambda_n } \, \, , \quad r_n := \sum_i \Phi^{(n)}_i
   .
   \label{eee}
\eeq
In the strong coupling limit where $1/f_0 \approx 0$ only configurations
with a vanishing eigenvalue of $G$ have a finite imaginary part.

We see from a simple electrodynamical analog that
for 2 and 3 centers $G$ can not
have vanishing eigenvalues. A zero eigenvalue is equivalent to
the existence of a nontrivial solution of the equation
\[
   V(\vec{r}_j ) := \sum_{i\ne j}
   \frac{1}{|\vec{r}_j - \vec{r}_i|} q_i = 0 .
\]
This is the condition for placing $A$ charges $q_i$ such
that the  electrostatic potential $V$ at the position of every
charge vanishes.  Obviously there is no solution
for less than four charges unless one charge is
removed to infinity.
If the number of scattering centers exceeds $3$
there is a continous
distribution of eigenvalues around zero in the
configuration ensemble. Therefore there are
configurations in  the ensemble which can produce scattering
resonances at values of $f_0$ much larger than the geometrical
length scale given by the internucleon distance.
This loss of the connection between the internal length scale given by
the eigenvalues of $G$ and the geometrical length scale introduced
by the density is the central result of the fixed scatterer calculation.
Note also that the observation of a critical particle number, trapping and the
loss of the geometrical length scale is very reminiscent of observations
made in the investigation of classical irregular scattering \cite{Smilansky}.

As we have seen, a nonvanishing eigenvalue distribution around zero is
responsible for the formation of a finite imaginary part of the scattering
amplitude in the stochastic model.
In figure \ref{eigen} the eigenvalue distribution for $\rm ^4He$
is shown. Its gross features can be understood again in the mean field
picture. The mean field matrix
\[
   \tilde{G}_{ij} = \frac{1}{R_0} (1 - \delta_{ij} )
\]
has one eigenvalue $\lambda_0 = (A-1)/R_0 $ and a $(A-1)$-fold degenerate
eigenvalue $\lambda_1 = - 1/R_0$. Only $\lambda_0$ contributes
to the scattering amplitude as its eigenvector
$ \Phi^{(0)} = (1,1, ... ,1)$ is the only one which has nonvanishing
overlap with the incoming wave $\exp(i\vec k\vec r_i)$ at $k=0$.
The right peak in figure \ref{eigen}
corresponds to $\lambda_0 $, the left peak to $\lambda_1$. The
smearing of the left peak due to fluctuations around the mean field matrix
causes zero eigenvalues to appear. Note that due to fluctuations no
qualitative change may occur since the existence of an eigenvalue $\lambda_0$
and the properties of the correspoding eigenvector are guaranteed by the
Perron--Frobenius theorem \cite{Graham}. We did not yet succeed finding an
analytical model for the small eigenvalues. From the point of view of
random matrix theory a theoretical calculation of the
eigenvalue distribution seems extremly difficult as the matrix entries
are highly correlated. On the other hand these correlations are found
to be essential since
the independence of the imaginary part of $F$ on $A$ is closely related to
correlations of the matrix entries.
If the matrix entries of $G$ are sampled inpedendently, a
decreasing density of small eigenvalues is found when $A$ is increased
whereas the correlated
matrices produce a stable value \cite{Dipl}.

\section{Inelastic cross sections}
\label{incl}
We will demonstrate
in this section that trapping of the projectile has a strong impact on the
behaviour of total and inelastic cross sections.
In order to show that our observation of a finite imaginary part
of the scattering amplitude is not connected with the limit of
zero energy scattering or large imaginary parts of the elementary
scattering amplitude let us consider a purely elastic elementary
interaction; i.e. annihilation processes like in
$\rm\bar p p$-scattering are absent. In this case all inelasticities
stem from transitions to excited states of the nucleus. To achieve this we
choose the s--wave amplitude to be unitary and use a simple parametrization
of the phase shift $\delta_0$ in terms of a real scattering length $a$
\beq
    f_0 = \frac{e^{2 i \delta_0} - 1}{2 i k};\qquad     \delta_0 = k a\quad
    \label{eq:unitary}
  .
\eeq
With this choice of $f_0$ the projectile-nucleon cross section
$\sigma_0 $ becomes constant in the limit $k\rightarrow 0$ and equal to
$\sigma_{0}= 4 \pi a^2$.
The total cross section of the projectile-nucleus scattering
amplitude is calculated via the optical theorem
and the inelastic cross section by subtracting the elastic cross section.
\beqa
   \sigma_{tot} & = & \frac{ 4 \pi}{k} \Im \langle F(\theta=0) \rangle \\
   \sigma_{inelast} & = & \sigma_{tot} - \sigma_{elast}.
   \label{eq:inn}
\eeqa
For one individual configuration the imaginary part of the
scattering amplitude $F$ vanishes like $k$ if the elementary
amplitude (\ref{eq:unitary}) is used, because
$F$ is the solution of a potential scattering problem with
a real potential. This statement is independent
of the number of scattering centers or the value of $a$.
The inelastic cross section for one configuration is zero,
because in potential scattering
total and elastic cross section are identical.
To see the origin of inelaticities
in the fixed scatterer model
we rewrite
equation (\ref{eq:inn}) in the following way:
\beqa
  \sigma_{inelast}
   & = & \frac{4\pi}{k} \Im \langle F(\theta=0) \rangle -
          \int d\Omega | \langle F (\Omega) \rangle |^2  \\
   & = &  \int d\Omega \langle  | F(\Omega) |^2 \rangle -
          \int d\Omega | \langle F (\Omega) \rangle |^2
\eeqa
The second line was obtained by using that averaging over
configurations is a linear and real operation and that
the optical theorem for one configuration yields
the elastic cross section. Inelastic processes
in the fixed scatterer model are thus related to fluctuations
of the scattering amplitude for the individual configurations.

We plot the cross sections as functions of $k$ for a scattering length
of $a=10 \rm \, fm$ in figures \ref{3incl} (three scattering centers)
and \ref{4incl} ($^4 \rm He$ ). For both calculations
CM corrected harmonic oscillator densities have been used \cite{Lenz}.
The width of the three center
density was fitted to give the same elastic cross
section as $^4 \rm He$. In the case of
$^4 \rm He$, total and inelastic cross section rise like
$1/k$ since the imaginary part of the scattering amplitude becomes constant
and equal to the value obtained in the strong coupling limit at zero energy.
Scattering in this example is dominated by inelastic processes.
The results are again found to be independent
of the scattering length $a$ once it is chosen large enough.
This is in contrast to the three center case where the
total cross section becomes constant at small values of $k$,
which means that the imaginary part of the scattering amplitude
is proprotional to $k$.

Obviously the existence of a strong coupling limit does not
depend on the existence of projectile-nucleon inelastic scattering.
The rise of the inelastic cross section is entirely due to excitation of
the nucleus. Although in the closure approximation all states are
degenerate, this fact alone is not sufficient to produce a rising inelastic
cross section as the three center calculation shows. The reason is that
despite the degeneracy transitions at low projectile momentum are
suppressed by the orthogonality of the nuclear wave functions which requires
a finite momentum transfer to be overcome. Therefore  it may be concluded
that it is rather the existence of resonant configurations which cause
this rise in the inelastic cross section in spite of decreasing momentum.
This can again be understood on the level of one configuration.
By an expansion of the solution of (\ref{a26}) for small $k$ one can see
that if the real part of the denominator in (\ref{eee}) becomes small,
the configuration almost developes a zero energy bound state.
This means that at a given small value of the energy the imaginary part
of the scattering amplitude can be very large although it ultimately
decreases
like $k$. The finite imaginary part of the averaged scattering
amplitude $\langle F \rangle$ is build up by these resonant
configurations.

For scattering
lengths which are comparable with the typical internucleon distances
a rising inelastic cross section is also observed for less
than four centers.
The regime of $a$ values
where rising inelastic cross sections
appear can be interpreted as the typical resonance length scale of
the system. The result of the fixed scatterer calculation can therefore be
restated in the following way: More than three target particles
give rise to resonant behaviour for arbitrarily large
values of $a$. Therefore the geometrical length scale given by the
internucleon distance is no longer connected to the resonant
length scale.

\section{Random potential calculation}
\label{Rand}
In the previous section
we identified
configurations with almost zero energy bound
states
as the origin
of large inelastic cross sections.
Nevertheless,
averaging over $k=0$ bound states is not sufficient to produce
a strong coupling limit. We demonstrate this by comparing our
results to scattering
off a random potential in the strong coupling limit.
Consider an attractive square well potential
\[
    V_R(r) = \alpha^2 (\Theta ( |r| - R)  - 1 )
\]
of range $R$ which we regard as a random variable with normalized
distribution $\rho(R)$. For infinitesimal small $k$ the scattering
amplitude is (see e.g. \cite{Merzbacher})
\[
   F = \frac{1}{\alpha {\rm \, cotan \,} \alpha R - i k}  - R ,
\]
and the condition to find a bound state at $k=0$ becomes
\[
  {\rm \, cotan \, } \alpha R = 0 \,\,  \Rightarrow  \,\,
  \alpha R = \frac{\pi}{2} (2 n +1) \, , \, n=0,1,...\,\, .
\]
For large $\alpha$ these resonant values of $R$ lie very dense
and averaging over a finite range of $R$ values always means
that many resonances are included. The average imaginary part
is then
\[
   \langle  \Im F \rangle  = \pi \int dR \rho(R) \delta
             ( \alpha {\rm \, cotan \,} \alpha R )
             = \frac{\pi}{\alpha^2} \sum_{n=0}^\infty
             \rho(\frac{\pi}{2\alpha} ( 2n + 1))           ,
\]
where we used $\lim_{\epsilon \rightarrow 0} \Im (x - i \epsilon)^{-1}
= \pi \delta(x)$.
In the limit $\alpha \rightarrow \infty$ the sum can be converted to
an integral as $\pi/\alpha$ is small yielding
\[
  \langle  \Im F \rangle  =
  \frac{1}{\alpha} \int dx \rho(x) = \frac{1}{\alpha}
{}.
\]
The imaginary part of the average scattering amplitude vanishes
like $1/\alpha$. This surprising result can be understood
by studying the wavefunctions inside the potential.
They are rapidly oscillating for large $\alpha$ and therefore
the overlap with the incoming wave decreases with increasing $\alpha$.
For the fixed scatterer model, the numerical results indicate an entirely
different behaviour \cite{Dipl}.
The overlap of the incoming wave with the eigenvectors of small eigenvalues
of $G$ does not depend strongly on the size of the eigenvalue when weighted
with their probability density. Thus
we can find resonances at arbitrarily large coupling and with nonvanishing
overlap of wavefunction and incoming wave.
Obviously a minimal criterion for a potential model to reproduce this
feature is that a decoupling between the scattering amplitude $f_0$ and
the strength of the potential must occur. So far we have not been able
to find a consistent model which would fulfill all requirements.

\section{Summary and Conclusion}
\label{concl}
In this work we investigated in comparison to the first order optical potential
a model in which in a simple way the many body aspect of multiple
scattering is partly retained.
We showed that as a consequence of the correlations occuring in the
multiple scattering process a strong coupling limit is found.
In terms of physical observables our
result of a finite imaginary part of the forward scattering amplitude
means that the total cross section increases like $1/k$ for small values
of $k$. As the elastic cross section remains finite this implies
that inelastic processes dominate low energy scattering for
four and more nucleons.
Our explanation for this is that, while trapped, the projetile travels for
a long time inside the nucleus thus having the chance of exciting arbitrary
nuclear states. The trapping mechanism causes a new resonant length scale
to emerge which is independent of the geometrical length scale and is
responsible for the non--suppression of the projectile wave in regions of
high target density which, however, could not be discussed in this work.

To formulate the fixed scatterer model we had to make approximations
which may be regarded as unrealistic in usual nuclear physics situations.
Therefore a direct application of our work to nuclear reaction data
seems not to be possible. However, the existence of trapping in
this model indicates that similar phenomena may be observable
under eralistic circumstances.
For example, the divergence of the total cross section
at zero energy is clearly an artefact of the closure approximation
since exciting the nucleus costs no energy.
Nevertheless a trapping mechanism may increase the probability
of inelastic processes, even
if energy loss is implemented. This will be subject to further
investigations.

Another interesting feature is the  particle number dependence of
the scattering amplitude. The fact that we find trapping
only for four and more centers shows that earlier work where only low
order backscattering effects were included in an improved optical potential
may still have serious drawbacks.
Systematic studies of the solutions, eigenvectors
and eigenvalues of the system of equations which determines the scattering
solutions for fixed target centers show that always a large
number of centers contribute to the trapping process \cite{Dipl}.
For this reason we failed to develop analytical models for our
numerical results by decomposing
one configuration into subclusters which trap the projectile.
Perturbative approaches
which take into account only a subclass of
backscattering processes
have little chance to describe the correct physics.

Further investigations are needed for finding models which can
reproduce the observed phenomena in a simple way.
Once this has been achieved
one may be able to find the relation of our results to seemingly
similar phenomena which are known as Anderson localization and classical
irregular scattering.

\vspace{1cm}\begin{center}{\bf Acknowledgments}\end{center}
We would like to thank J.--P.Dedonder, F.Lenz and M.Thies
for discussions, K.Yazaki for comments on the manuscript and D.Lehmann
and E.Schneid for communicating their results to us prior to publication.
We would also like to acknowledge the kind hospitality of the
Universit\'e Paris 7
where part of this work has been carried out.

\newpage
\begin{figure}
\caption{Zero energy scattering amplitude for $^4\rm He$
as function of $|f_0|$
for various arguments $\varphi$ of $f_0$.
The full line is the real part, dotted line imaginary part.}
\label{Real4}
\end{figure}
\begin{figure}
\caption{Ratio of imaginary to real part of the zero energy
scattering amplitude for $^{16}\rm O$ as function of $|f_0|$
for various arguments $\varphi$ of $f_0$.
The full line is the stochastic calculation, dashed line the
optical potential.}
\label{O16}
\end{figure}
\begin{figure}
\caption{Zero energy scattering amplitude as function of
the number of scattering centers $A$.
A homogenous density has been used,
$f_0 = 250.0{\rm \, fm}+ i \times 6.3{\rm \, fm}$. The full curve is the
prediction of the mean field calculation, the points are the
results of the stochastic calculation (squares: real part; triangles:
imaginary part).}
\label{Adependence}
\end{figure}
\begin{figure}
\caption{Eigenvalue distribution of $G$
(eq.\protect\ref{gij})
for $ ^4\rm He$.
The distribution is normalized to $1$,
therefore it is the probability density
of  finding a given eigenvalue in a randomly picked configuration.
The plot is taken from Dirk Lehmanns
diploma thesis\protect\cite{Dipl}.}
\label{eigen}
\end{figure}
\begin{figure}
\caption{Cross sections for 3 scattering centers as function of projectile
  momentum $k$. The scattering length is $a=10 \rm \, fm$.
  The full curve is the total cross section, dotted curve
  elastic cross section, dashed curve inelastic cross section.}
\label{3incl}
\end{figure}
\begin{figure}
\caption{Cross sections for 4 scattering centers as function of projectile
  momentum $k$. The scattering length is $a=10 \rm \, fm$.
  The full curve is the total cross section, dotted curve
  elastic cross section, dashed curve inelastic cross section.}
\label{4incl}
\end{figure}
\end{document}